\documentstyle[15lomcon,cite,epsfig]{article}

\begin{document}

\title{New oscillation results from the T2K experiment}

\author{ Alexander Izmaylov \footnote{e-mail: izmaylov@inr.ru} \\
(for T2K Collaboration)}

\address{Institute for Nuclear Research RAS, 117312 Moscow, Russia}


\maketitle\abstracts{ The T2K (Tokai to Kamioka) experiment is a second generation long baseline neutrino
oscillation experiment located in Japan. The main goal is to probe the $\theta_{13}$ neutrino mixing parameter by looking for $\nu_{\mu}\to\nu_e$ transitions in an almost pure beam of muon neutrinos. The T2K utilizes the neutirno beam produced at J-PARC (Tokai, Ibaraki) and Super-Kamiokande (Kamioka, Gifu) is used  as a far detector. The experiment has been in operation since January 2010. 
After analyzing 1.43$\times$10$^{20}$ p.o.t.  data collected  six events are observed in far detector while the expected number with sin$^2 2\theta_{13}$=0 is 1.5$\pm$0.3. Null oscillation hypotheis leads to 7$\times$10$^{-3}$ probability to observe six or more candidate events, which so gives 2.5 $\sigma$ significance to the result. Thus the current T2K result is an indication of $\nu_e$ appearance due to $\nu_{\mu}\to\nu_e$ transitions. As for the first T2K $\nu_{\mu}$ disappearance data, the null oscillation hypothesis is exluded at 4.5 $\sigma$ level and the estimated atmospheric mixing parameters are consistent with the results from Super-Kamiokande and MINOS experiments.}
\section{Introduction}
The phenomenon of neutrino oscillations established and confirmed by a number of challenging experiments in the past 20 years appears to be quite compelling. The goal for the experimentalists now is to explore the case further. 

The T2K (Tokai to Kamioka) \cite {t2k_overview} experiment is a second generation long baseline (LBL) neutrino
oscillation experiment. The main goal is to probe the only unknown $\theta_{13}$ neutrino mixing parameter by looking for $\nu_{\mu}\to\nu_e$ transitions in an almost pure beam of muon neutrinos. Non-zero $\theta_{13}$ is crucial for furhter experimental searches for CP-violation in lepton sector.  The best upper limit of $\sin^22\theta_{13}<0.15$ (90\% C.L.) was obtained by CHOOZ reactor experiment \cite{chooz_nue} (1999) and further slightly corrected by LBL MINOS \cite{minos_2010_nue} (2010).  

Precise measurement of the atmospheric $\Delta m^2_{23}$ and of the $\theta_{23}$ mixing parameters in $\nu_{\mu}$ disappearance channel forms another goal of the experiment.

The T2K is an international collaboration which includes about 500 members from 58 institutes of 12 countries.
\section{T2K design}
\begin{figure}[t]
\centering
\includegraphics[width=0.8\textwidth ]{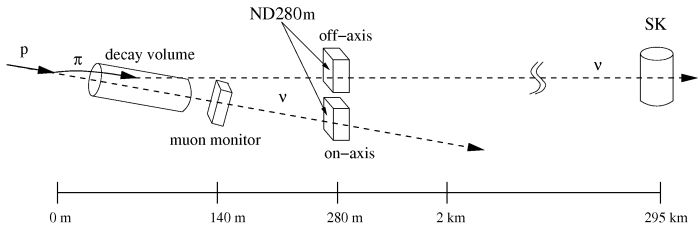}
\caption{A schematic view of the T2K LBL exepriment.}
\label{fig:layout}
\end{figure}

 T2K neutrino beam is produced using 30 GeV proton synchrotron at the Japan  Proton Accelerator Research Complex (J-PARC) in Tokai, Japan. The experiment layout is presented on fig.~\ref{fig:layout}.

 The T2K is the first neutrino experiment to utilize an  "off-axis" neutrino beam conception. The neutrino beam at J-PARC is designed so that it is directed 2.5$^{\circ}$ away from the direction of the Super-Kamiokande (Super-K), located 295 km to the west. This results in forming a narrow energy band neutrino beam peaked at about 600 MeV. A peak is put on the oscillation maximum for the atmospheric $\Delta m^2$ scale at the Super-K site. At the same time the usage of the "off-axis" design helps to minimize the background for $\nu_e$ searches. The neutrino beam is monitored by a set of detectors at J-PARC: non-magnetized on-axis detector INGRID  and magnetic off-axis near detector, both located at 280 m from the proton target and forming an ND280 detector complex. The main goal of the complex is to provide neutrino spectra and beam flavor composition measurements prior to oscillation process as well as to measure neutrino interactions cross-sections. 50 kton Super-K massive water Cherenkov detector is used as a T2K far detector. The Super-K performance is well-matched at sub-GeV level and provides good $\mu-$ and $e-$ (``fuzzy") single rings separation with about 99\% efficiency.   

   Construction of the J-PARC neutrino beamline was started in April 2004. The complete chain of
accelerator and neutrino beamline was successfully commissioned during 2009. The T2K started physics data taking in January 2010.

\section{T2K neutrino oscillation analysis}
The T2K data was collected in two runs RunI (Jan--Jun 2010) and RunII (Nov 2010--Mar 2011). 145 kW stable beam operation was achieved in RunII but the run was stopped in March 2011 due to the Great East Japan Earthquake.  The total data set used in the present $\nu_e$ appearance \cite{t2k_result} and $\nu_{\mu}$ disappearance analysis corresponds to 1.43$\times10^{20}$ p.o.t. (2\% of the T2K final goal).  The beam direction stability which is a key point for an off-axis beam was checked to be well within 1 mrad (this corresponds to neutrino peak energy shift of 2\%) during all the data taking. 

At T2K energies CCQE interactions are dominant and so corresponding leptons form Super-K $\nu_{\mu}$ and $\nu_e$ signals. For $\nu_{\mu}\to\nu_e$ search two main background sources are  intrinsic $\nu_e$ contamination in $\nu_{\mu}$ beam and $\pi^{0}$s from  NC interactions. As for $\nu_{\mu}$ disappearance channel CC1$\pi$ interactions form the main background.
Super-K event selection criteria were optimized in order to suppress background sources. It is worth mentioning that all the selection cuts in the far detector were predefined and fixed based on Monte-Carlo (MC) simulation results prior to analysis so to avoid any biases. 

In order to evaluate oscillation parameters the number of events that pass Super-K selection cuts $N^{obs}_{SK}$ is compared  
with the calculated expectation $N^{exp}_{SK}$ which is based on the predicted neutrino flux, external cross-section data and $\nu_{\mu}$ CC inclusive $R_{\frac{Data}{MC}}$ rate measurement in the off-axis near detector. The latter one is used as a normalization factor. The present analysis is based on NEUT \cite{neut} as a neutrino interactions generator. As for the neutrino flux prediction it uses the information from proton beam monitors as an input for further FLUKA simulation. The results are then tuned to the p+C measurements from NA61/SHINE CERN experiment \cite{na61}.  For some cases (horn focusing, out of target interactions) GEANT3 (GCALOR) MC is also utilized with the cross-sections tuned to external data. 

General $\nu_{\mu}$ and $\nu_{e}$ CCQE event selection cuts used in the T2K far detector are as follows: the event is accepted when its time is within -2 -- +10 $\mu$sec beam trigger on-time window, it is a fully-contained (FC) event (Super-K is divided into two detectors: Outer detector (OD) and Inner detector (ID), the cut requires the vertex to be inside the ID and no activity in the OD), we then proceed with FCFV events which have vertexes inside 22.5 kton fiducial volume (FV, the vertex should be $>$ 200 cm from the nearest ID wall),  among the selected we  deal with single ring events to which Super-K PID algorithms based on ring shape and opening angle are further applied. 41 total Super-K events were selected using the above criteria: 33 $\mu-$ and 8 $e-$like.

\begin{figure}[t]
\centering
\includegraphics[width=0.45\textwidth ]{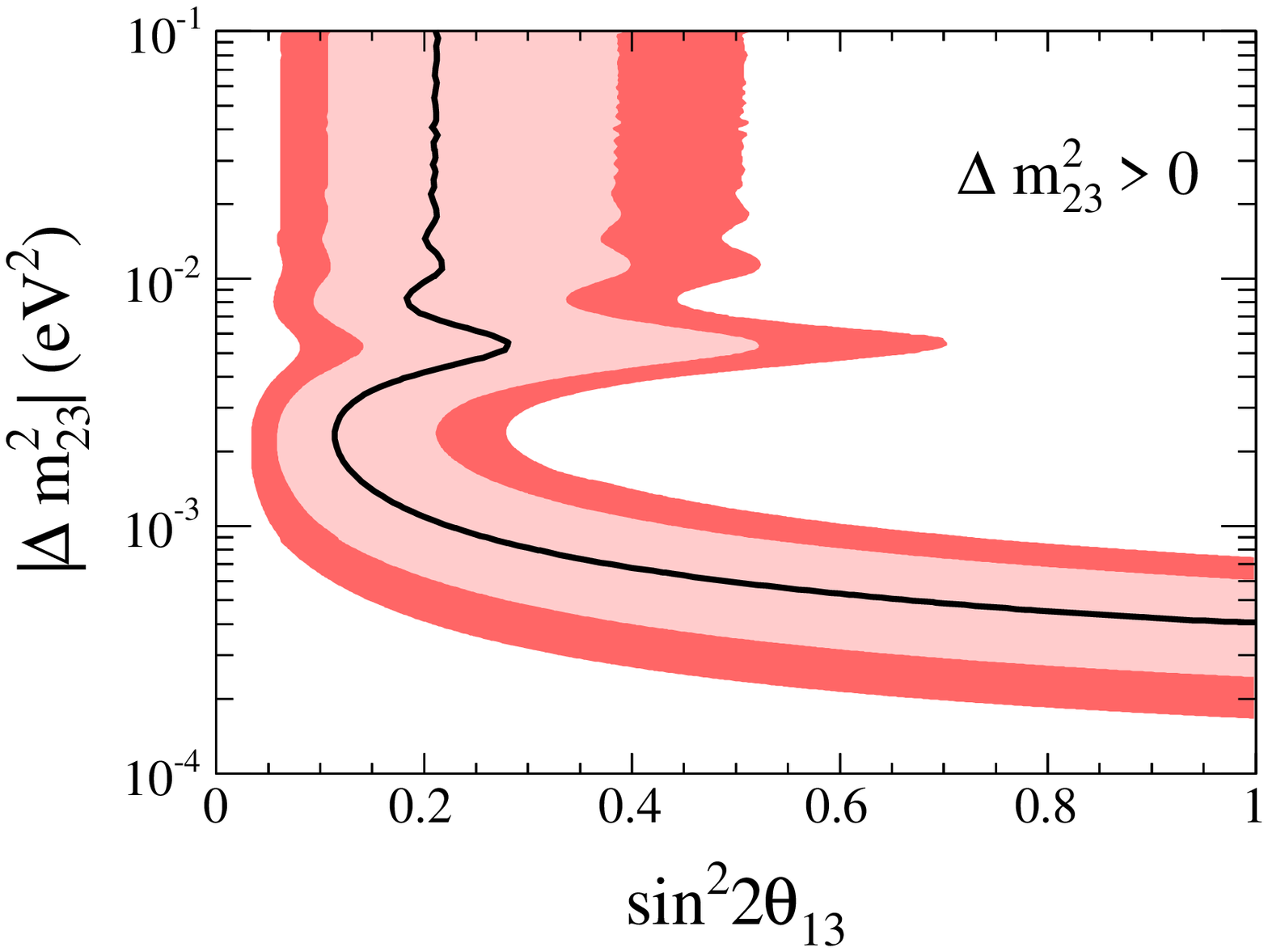}
\includegraphics[width=0.45\textwidth ]{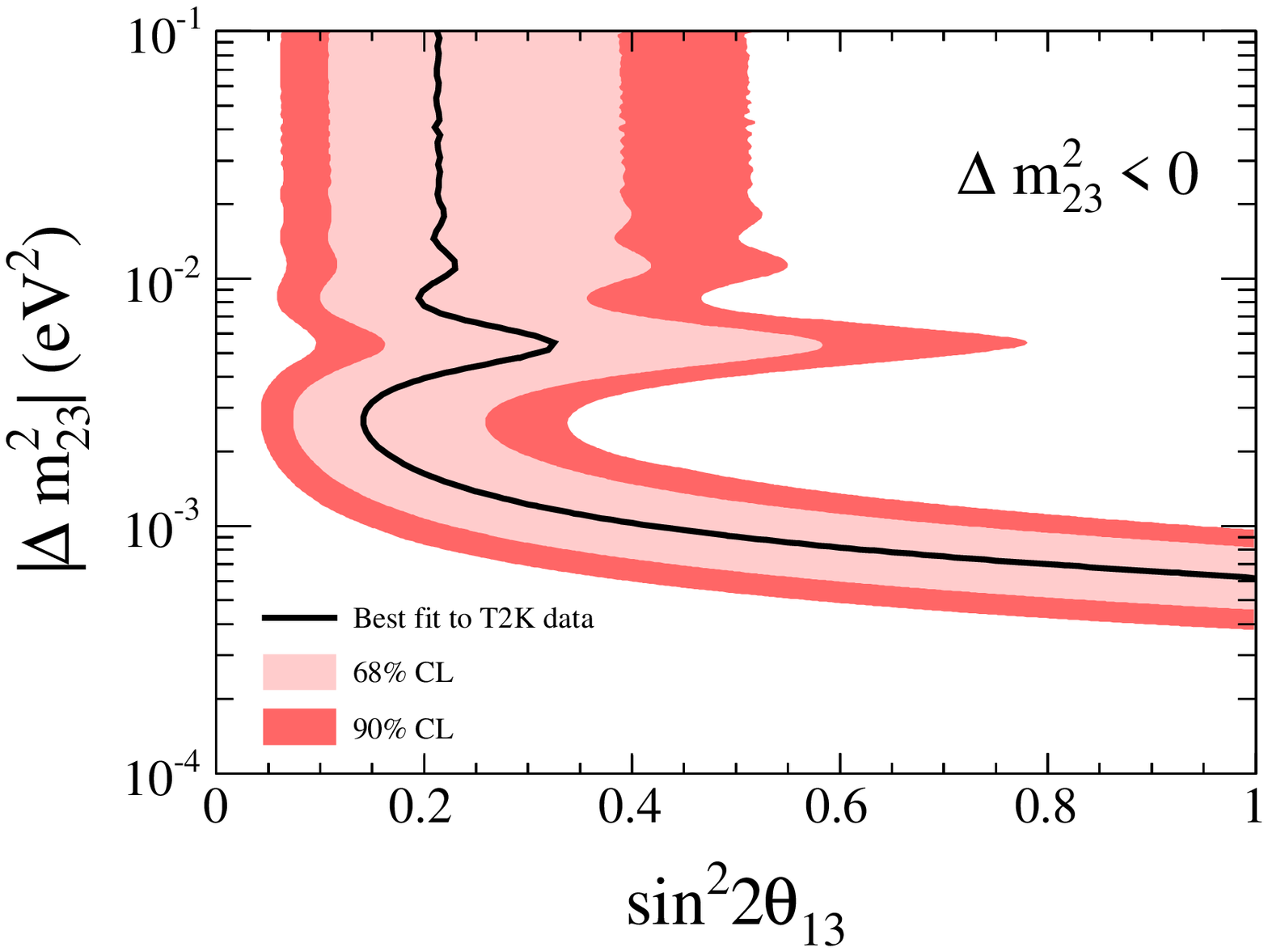}
\caption{$\sin^22\theta_{13}$--$|\Delta m^2_{23}|$ contours for $\nu_e$ appearance. Best fit, 68\% and 90\% C.L. regions are shown for normal (left) and inverted (right) mass hierarchy. $\delta_{CP}$ = 0.}
\label{fig:nue}
\end{figure}
 
\section{$\nu_e$ appearance results}
The $\nu_{\mu}\to\nu_e$ appearance search is started with 8 $e$-like events remained after ``basic" selection criteria. The further cuts are optimized to minimize intrinsic $\nu_e$ and NC$\pi^0$ backgrounds, the cuts are: visible energy $E_{vis}$ should be $>$ 100 MeV (7 events survived), no Michel electrons (6 events), $M_{inv}<105$  MeV (calculated under two $e$-like rings assumption), reconstructed neutrino energy $E^{\nu}_{rec}<1250$ MeV (all 6 events passed two last cuts). The number of $\nu_e$ candidates is therefore $N^{obs}_{SK}$=6. The  expectation gives (backgrounds + $\nu_{\mu}\to\nu_e$ Solar term) $N^{exp}_{SK}=1.5\pm 0.3$ for zero $\theta_{13}$ taking into account $^{+22.8\%}_{-22.7\%}$ systematics. The probability to observe six or more events is then 0.7\% which corresponds to 2.5~$\sigma$ significance. The Feldman-Cousins \cite{feld_cous} method was implemented to get confidence intervals for oscillation parameters:  0.03(0.04)$<$ sin$^2 2\theta_{13}$ $<$0.28(0.34) at 90\% C.L. for $\delta_{CP}$ = 0, $\sin^22\theta_{23}=1.0$, $|\Delta m^2_{23}|=2.4\times10^{-3}$ eV$^2$ and normal (inverted) mass hierarchy (fig.~\ref{fig:nue}), the corresponding central value is 0.11(0.14).

\section{$\nu_{\mu}$ disappearance results}
\begin{figure}[t]
\centering
\includegraphics[width=0.45\textwidth ]{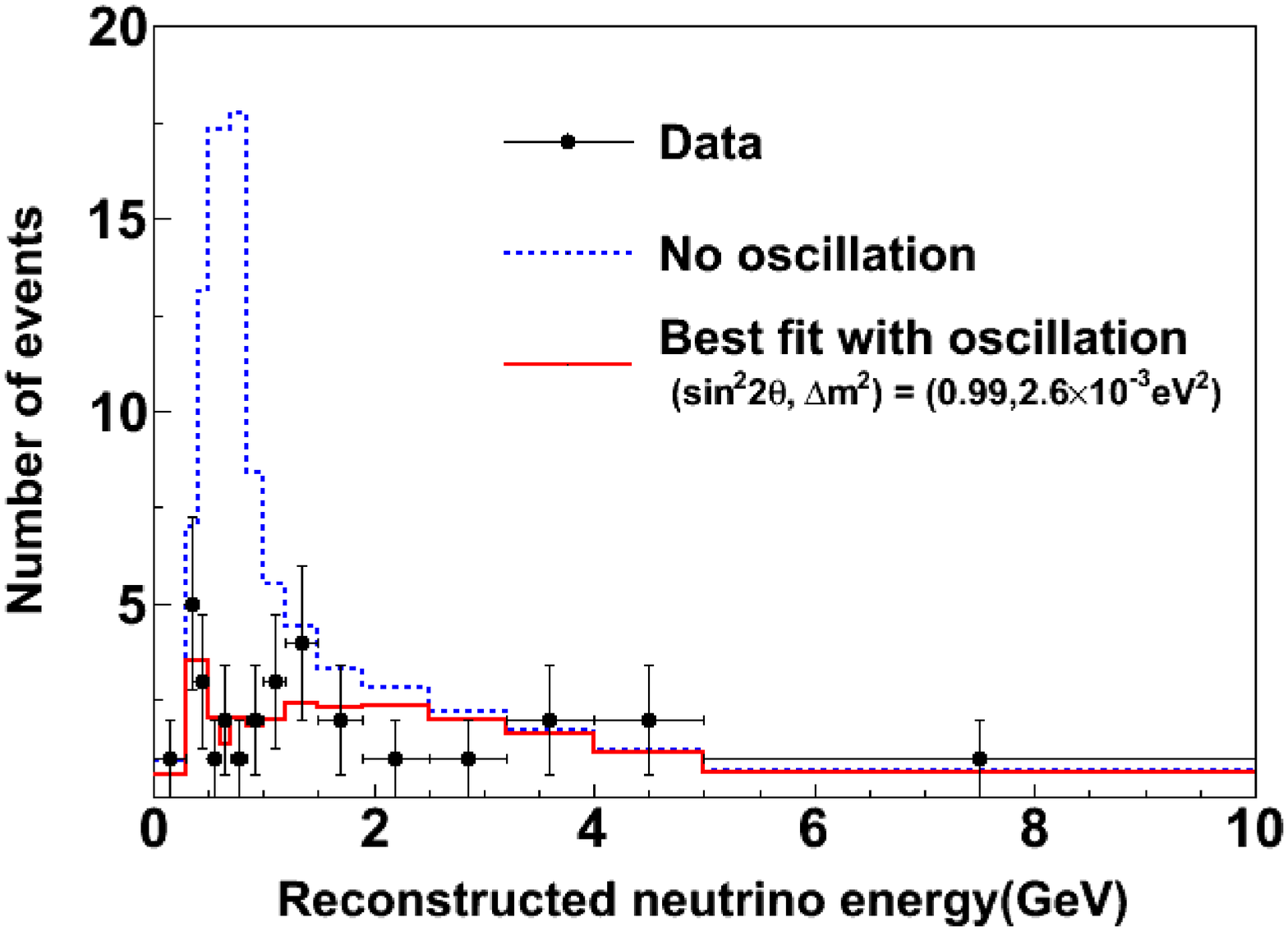}
\includegraphics[width=0.45\textwidth ]{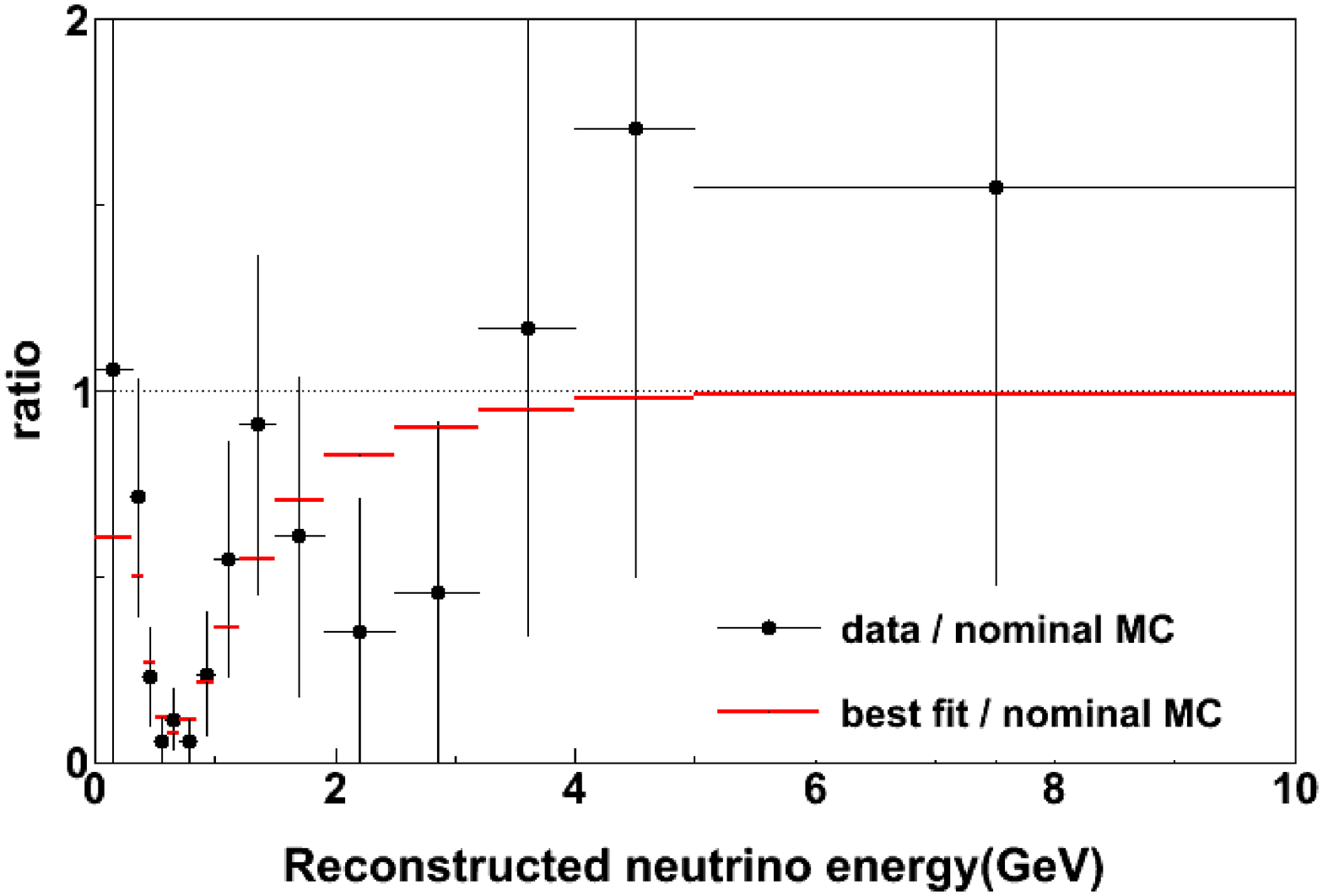}
\caption{Reconstructed neutrino energy at Super-K for selected $\nu_{\mu}$ events (left) and Data/MC spectra ratio (right).}
\label{fig:numu}
\end{figure}

For $\nu_{\mu}$ disappearance analysis ``basic" Super-K cuts are followed by two additional ones: require less than two decay electrons and reconstructed muon momentum $P^{\mu}_{rec}>$200 MeV. 31 events survived all cuts. The null oscillation hypothesis gives 104 events with  $^{+13.2\%}_{-12.7\%}$ systematics, it is therefore exluded at 4.5 $\sigma$ level.  The energy spectrum distortion due to oscillations can be also clearly observed (fig.~\ref{fig:numu}). The combined analysis based on events number and energy spectrum shape allows to set more stronger limit on null oscillation hypothesis, the probability of null oscillation is limited to $\sim10^{-10}$.  
 Mixing parameters were retrieved using Feldman-Cousins algorithm and two-flavor oscillation model:
$\sin^22\theta_{23}>0.85$ and $2.1\times 10^{-3}<\Delta m^2_{23}<3.1\times 10^{-3}$ eV$^2$ (90\% C.L.). The obtained values are consistent with the previous measurements by Super-Kamikande and MINOS \cite{sk_atm,minos_atm}.
\section{Conslusions}  
Neutrino oscillation results from the T2K second generation long baseline experiment are presented. The T2K is the first LBL experiment to utilize off-axis conception of neutrino beam. The present analysis is based on 1.43$\times 10^{20}$ p.o.t. collected from January 2010 to March 2011 (2\% of the T2K final goal).  Total six $e$-like single-ring events passed all the selection criteria.  The probability to observe six or more events under zero $\theta_{13}$ hypothesis is 0.7\% which is equivalent to 2.5 $\sigma$ significance. The corresponding 90\% C.L.
intervals are  0.03(0.04)$<$ sin$^2 2\theta_{13}$ $<$0.28(0.34) at 90\% C.L. for $\delta_{CP}$ = 0 and normal (inverted) mass hierarchy. Hence the T2K result is an indictaion of non-zero $\theta_{13}$.

As for $\nu_{\mu}$ disappearance analysis the null oscillation hypothesis is excluded at 4.5 $\sigma$ level. The obtained oscillation parameters are consistent with the results from Super-Kamiokande and MINOS experiments.   

The restart of J-PARC accelerator operation is planned for December 2011 and T2K is on schedule to resume physics data taking in early 2012. 

\section*{Acknowledgments}

   The present work was supported in part by the ``Neutrino Physics" Program of the Russian Academy of Sciences, by the RFBR (Russia)/JSPS
(Japan) grant ¹ 11-02-92106 and by the Science School grant ¹ 65038.2010.2.

\end{document}